\documentclass[10pt]{iopart}
\usepackage{graphicx}
\usepackage{color}

\expandafter\let\csname equation*\endcsname\relax

\expandafter\let\csname endequation*\endcsname\relax

\usepackage[libertine]{newtxmath}
\usepackage{slashed}
\usepackage{amsfonts}
\usepackage[normalem]{ulem}

\usepackage{microtype}\usepackage[colorlinks,linkcolor=blue,anchorcolor=blue,citecolor=blue,urlcolor=blue,breaklinks=true]{hyperref}

\expandafter\let\csname equation*\endcsname\relax

\expandafter\let\csname endequation*\endcsname\relax

\def\be{\begin{equation}}
\def\ee{\end{equation}}
\def\bea{\begin{eqnarray}}
\def\eea{\end{eqnarray}}

\begin{document}

%\preprint{ADP-24-05/T1244}

\title{Challenges in the extraction of physics beyond the Standard Model from electron scattering}

\author{Xuan-Gong Wang$^1$\footnote{To whom correspondence should be addressed.}, A W Thomas$^1$}
\address{$^1$ ARC Centre of Excellence for Dark Matter Particle Physics and CSSM, Department of Physics, University of Adelaide, Adelaide SA 5005, Australia}
\eads{\mailto{xuan-gong.wang@adelaide.edu.au}, \mailto{anthony.thomas@adelaide.edu.au}}

\begin{abstract}
{Precise measurements of electron and positron scattering, including parity violation, offer great promise in the search for physics beyond the Standard Model. In this context it is crucial to understand the corrections which might arise from charge symmetry violation, as well as the less well known strange and charm quark distributions. Our analysis, using state of the art parton distributions, suggests that these contributions lead to corrections in the extraction of the weak couplings $g^{eq}_{AV}$ and $g^{eq}_{VA}$ of the order $(1-2)\%$, while they are as large as $4\%$ for $g^{eq}_{AA}$, at a typical scale of $Q^2 = 10\ {\rm GeV}^2$. 
These results underline the importance of carrying out high precision measurements, which will not only provide information on physics beyond the Standard Model but also reduce the current uncertainties on our knowledge of the strange and charm quark distributions in the proton.}
\end{abstract}


\submitto{\jpg}

\maketitle

%===============================================================%
%===============================================================%
\section{Introduction}

Parity-violating electron scattering (PVES) has emerged as a powerful tool to test the Standard Model (SM)~\cite{Prescott:1978tm, Prescott:1979dh,Young:2007zs}, probe new physics~\cite{Thomas:2022qhj, Thomas:2022gib, Boughezal:2021kla, Crivellin:2021bkd} and study hadron and nuclear structure~\cite{Cloet:2012td, Qweak:2018tjf, PREX:2021umo, Liang:2023bid, Beminiwattha:2023est, Becker:2018ggl}.
In particular, considerable effort has gone into the preparation of deep inelastic PVES experiments aimed at accurately determining the fundamental couplings $g^{eq}_{AV}$, $g^{eq}_{VA}$, and $g^{eq}_{AA}$~\cite{Arrington:2021alx}. At leading order, these couplings are defined as the products of the lepton and quark weak couplings in the Standard Model~\cite{Erler:2013xha}, otherwise known as $C_{1q}$, $C_{2q}$, and $C_{3q}$.

At Jefferson Lab the PVDIS Collaboration~\cite{PVDIS:2014cmd, Wang:2014guo} extracted $2 C_{2u} - C_{2d}$ from measurements of the parity violating asymmetry in deep inelastic scattering of polarised electrons from a deuteron target, using the $C_{1q}$ determined from the proton weak charge and atomic parity violation data.
In the future, the SoLID Collaboration~\cite{Chen:2014psa, JeffersonLabSoLID:2022iod} aims to repeat the earlier PVES measurements with much higher precision and over a wider range of $x$ and $Q^2$. As a result it will improve the knowledge of $g_{VA}^{eq}$ by an order of magnitude.

In the analysis of such experiments, the role of the heavier strange and charm quark distributions requires careful attention. In addition, for deuterium targets, one needs to question the validity of charge symmetry~\cite{Londergan:2009kj} between the proton and neutron parton distribution functions (PDFs) ( i.e., $u^n = d^p$, and $d^n = u^p$ ).

Given advances in experimental capabilities in recent years, it is required that theoretical predictions are computed with high accuracy. Moreover, given that new physics effects are typically expected to be small, precise determinations of hypothesised new physics signals rely on refined analyses of the PVES data. 
In this paper, we systematically investigate heavy quark and charge symmetry violation (CSV) effects on PVES asymmetries with a deuteron target.
We show that, in order to maximise the impact of such experiments on new physics beyond the SM, one needs to have a much better knowledge of the strange and charm quark distributions in particular.

Charge symmetry between proton and neutron PDFs is expected to be violated~\cite{Londergan:2009kj} because of the mass difference between $u$ and $d$ quarks~\cite{Rodionov:1994cg,Sather:1991je,Close:1988br} and the electromagnetic interaction~\cite{Spiesberger:1994dm, Martin:2004dh, Wang:2015msk}. It is now well known that CSV gives rise to a sizeable correction~\cite{Londergan:2003ij,Bentz:2009yy} to the NuTeV measurement of $\sin^2\theta_W$~\cite{NuTeV:2001whx}.

The charge conjugation positive combinations $s^+ \equiv s + \bar{s}$ and $c^+ \equiv c + \bar{c}$ are included in recent global PDF analyses~\cite{Bailey:2020ooq, Hou:2019efy, Ball:2022qks, Harland-Lang:2024kvt} of deep inelastic data, although they are less well known than those of the $u$ and $d$ quarks. However, when it comes to the charge conjugation odd (C-odd) distributions, $s^- \equiv s_V = s - \bar{s}$ and $c^- \equiv c_V = c-\bar{c}$, the relative uncertainties in our present knowledge are considerably larger. These distributions can arise perturbatively at next-to-next-to-leading order (NNLO)~\cite{Catani:2004nc}. Their non-perturbative generation has been investigated in the meson-baryon cloud model~\cite{Holtmann:1996be,Melnitchouk:1996fj,Melnitchouk:1997ig, Hobbs:2013bia}, chiral effective theory~\cite{Wang:2016eoq, Wang:2016ndh, Salamu:2019dok}, and the light-front holographic model (LFHQCD)~\cite{Sufian:2018cpj, Sufian:2020coz}. While most global QCD analyses of PDFs have introduced asymmetric $s$ and $\bar{s}$ distributions~\cite{Bailey:2020ooq, Cocuzza:2021cbi, NNPDF:2021njg, Ablat:2024muy}, early analyses did not show unambiguous evidence for non-vanishing C-odd charm distributions~\cite{Guzzi:2022rca}. The first determination of a sizeable intrinsic $c-\bar{c}$ distribution from the global fit of PDFs was only reported recently by the NNPDF Collaboration~\cite{NNPDF:2023tyk}. The asymmetry, $c_V$, in their charm PDF is qualitatively in agreement with the result of meson-baryon cloud model~\cite{Hobbs:2013bia}, but opposite to that of the LFHQCD calculation~\cite{Sufian:2020coz}. They also reported an asymmetry, $s - \bar{s}$, in the strange quark  distribution, which is one order of magnitude larger than the non-perturbative result calculated in~\cite{Wang:2016eoq}.

 In section~\ref{sec:beam-asymmetry}, we briefly review the beam asymmetries in parity-violating deep inelastic scattering. In section~\ref{sec:corrections}, we derive the corrections to the beam asymmetries arising from CSV, as well as the strange and charm quark distributions. 
We present the numerical results in section~\ref{sec:results}, and discuss the impact on new physics searches in section~\ref{sec:new-physics}. A summary of our conclusions is reported in section~\ref{sec:conclusion}.

%%%%%%%%%%%%%%%%%%%%%%%%%%%%%%%%%%%%%%%
\section{Beam asymmetry}
\label{sec:beam-asymmetry}
In the Standard Model (SM), the tree level weak couplings to quarks and leptons are
\begin{eqnarray}
\label{eq:couplings}
\{ g^e_V, g^u_V, g^d_V\} &=& \{ - \frac{1}{2} + 2 \sin^2\theta_W\, , \frac{1}{2} - \frac{4}{3}\sin^2\theta_W \, , - \frac{1}{2} + \frac{2}{3}\sin^2\theta_W\}\, , \nonumber\\
\{ g^e_A, g^u_A, g^d_A\} &=& \{ - \frac{1}{2}\, , \frac{1}{2}\, , -\frac{1}{2} \} \, ,
\end{eqnarray}
 where $\theta_W$ is the weak mixing, or Weinberg, angle. We note that there are radiative corrections~\cite{Marciano:1982mm, Czarnecki:1995sz,Erler:2003yk} which must be included in a complete analysis of the data once it is taken. We do not show these effects because the corrections are well known and the effect of their uncertainties much smaller than those considered here. The products of the lepton and quark couplings are the quantities needed for testing the SM and searching for new physics~\cite{Zheng:2021hcf},
 \begin{equation}
 g^{eq}_{AV} = 2 g^e_A g^q_V\, ,\ \ 
 g^{eq}_{VA} = 2 g^e_V g^q_A\, ,\ \ 
 g^{eq}_{AA} = - 2 g^e_A g^q_A\, .
\end{equation}
We start from the double differential cross section given in~\cite{Thomas:2022qhj} (neglecting the dark photon contribution)
\begin{eqnarray}
\frac{d^2 \sigma}{dx dy}
&=& \frac{4\pi \alpha^2 s}{Q^4}
\Big(
[x y^2 F_1^{\gamma} + f_1(x,y) F_2^{\gamma}] 
 - \frac{1}{\sin^2 2\theta_W}\frac{Q^2}{Q^2 + M_Z^2} \times \nonumber\\ 
&& (g^e_V - \lambda g^e_A)  [x y^2 F_1^{\gamma Z} + f_1(x,y) F_2^{\gamma Z} - \lambda x y (1-\frac{y}{2}) F_3^{\gamma Z}]  \Big)\, ,
\end{eqnarray}
where  $f_1(x,y) = 1 - y - xyM/2E \approx 1 - y$ and $\lambda = + 1 (-1)$ represents positive (negative) initial electron helicity. 
 For positron scattering, the cross sections can be obtained with $g^e_A$ being replaced 
 by $-g^e_A$~\cite{Anselmino:1993tc}. In the DIS scheme, the structure functions  can be expressed in terms of PDFs as~\cite{Anselmino:1993tc}
 \bea
 F_1^{\gamma} &=& \frac{1}{2} \sum_q e^2_q (q + \bar{q})\, ,\ F_2^{\gamma} = 2 x F_1^{\gamma}\ ,\nonumber\\
 F_1^{\gamma Z} &=& \sum_q e_q g^q_V (q + \bar{q})\, ,\ F_2^{\gamma Z} = 2 x F_1^{\gamma Z}\, ,\nonumber\\
 F_3^{\gamma Z} &=& 2 \sum_q e_q g^q_A (q - \bar{q})\ ,
 \eea
where $e_q$ is the electric charge of the quark $q$.

The parity-violating asymmetry in electron scattering is defined by
\be
\label{eq:A_RL_electron}
A^{e^-}_{RL} = \frac{\sigma^{e^-}_R - \sigma^{e^-}_L}{\sigma^{e^-}_R + \sigma^{e^-}_L} 
= \frac{1}{\sin^2 2\theta_W} \frac{Q^2}{Q^2 + M^2_Z}
\Bigg[
\frac{g^e_A F_1^{\gamma Z}}{F_1^{\gamma}}
+ \frac{y(1-\frac{y}{2})}{1 + (1-y)^2} \frac{g^e_V F^{\gamma Z}_3}{F^{\gamma}_1}
\Big]\, . 
\ee
In addition, the difference between unpolarized electron and positron scattering can be written as
\be
\label{eq:A_electron_positron}
A^{e^+ e^-} = \frac{\sigma^{e^+} - \sigma^{e^-}}{\sigma^{e^+} + \sigma^{e^-}}
= \frac{1}{\sin^2 2\theta_W} \frac{Q^2}{Q^2 + M^2_Z} \frac{y (1- \frac{y}{2})}{1 + (1-y)^2} \frac{g^e_A F_3^{\gamma Z}}{F_1^{\gamma}}\, .
\ee
For $Q^2 \ll M_Z^2$, these asymmetries can be rewritten in terms of the Fermi constant $G_F= 1.1663787\times 10^{-5}\ {\rm GeV}^{-2}$ using the relation
\begin{equation}
\label{eq:G_F}
\frac{Q^2}{\sin^2 2 \theta_W(Q^2 + M_Z^2)} = \frac{G_F Q^2}{2\sqrt{2} \pi \alpha} \, .
\end{equation}
%

%%%%%%%%%%%%%%%%%%%%%%%%%%%%%%%%%%%
\section{Heavy quark and CSV corrections}
\label{sec:corrections}
For a deuteron target, taking into account charge symmetry violation (CSV)~\cite{Londergan:2009kj},
\be
\delta u = u^p - d^n\, ,\ \delta d = d^p - u^n\, ,
\ee
the structure functions can be expressed in terms of the proton PDFs as
\bea
F_1^{\gamma} = \frac{5}{18} (u^+ + d^+) + \frac{4}{9} c^+ + \frac{1}{9} s^+ 
+ \frac{1}{18} (- 4 \delta d^+ - \delta u^+)\, ,\nonumber\\
F_1^{\gamma Z} = \frac{1}{3} (2 g^u_V - g^d_V) (u^+ + d^+)
+ \frac{1}{3} (- 2 g^u_V \delta d^+ + g^d_V \delta u^+) 
+ \frac{4}{3} g^u_V c^+ - \frac{2}{3} g^d_V s^+\, ,\nonumber\\
F_3^{\gamma Z} 
= \frac{2}{3} (2 g^u_A - g^d_A) (u_V + d_V)
+ \frac{2}{3} (-2 g^u_A \delta d_V + g^d_A \delta u_V) 
+ \frac{8}{3} g^u_A c_V - \frac{4}{3} g^d_A s_V\, .\nonumber\\
\eea
In these above expressions, we have assumed $c^n_V = c^p_V$, $s^n_V = s^p_V$, and $c^+_n = c^+_p$, $s^+_n = s^+_p$.

Following the notation in~\cite{Zheng:2021hcf}, one can define
\bea
Y(y) &=& \frac{1-(1-y)^2}{1+(1-y)^2}\, , \ \ R_V = \frac{u_V + d_V}{u^+ + d^+}\, ,\nonumber\\
R_C &=& \frac{2(c+ \bar{c})}{u^+ + d^+}\, ,\ \ \ \ \ \ \  R_S = \frac{2(s+ \bar{s})}{u^+ + d^+}\, .
\eea
%

%%%%%%%%%%%%%%%%%%%%%%%%%%%%%%%%%%%%%%%%%%%%%%%%%%%%%%%%%%%%
\subsection{$A^{e^-}_{RL,d}$}

The asymmetry in~(\ref{eq:A_RL_electron}) then becomes
\be
\label{eq:A_RL_electron_2}
A^{e^-}_{RL,d} = \frac{3 G_F Q^2}{10 \sqrt{2} \pi \alpha} 
\Big[ (2 g^{eu}_{AV} - g^{ed}_{AV}) (1 + \Delta_{AV}) + R_V Y (2 g^{eu}_{VA} - g^{ed}_{VA}) (1 + \Delta_{VA})\Big]\, ,
\ee
where $\Delta_{AV}$ and $\Delta_{VA}$ are correction factors arising from the CSV, heavier $s$ and $c$ quark distributions
\be
\label{eq:Delta_AV_VA}
\Delta_{AV(VA)} = \Delta^{\rm CSV}_{AV (VA)} + \Delta^s_{AV (VA)} + \Delta^c_{AV (VA)}\, .
\ee
The individual contributions read
\bea
\label{eq:Delta-AV}
\Delta^{\rm CSV}_{AV} &=& \frac{2 (g^u_V + 2 g^d_V) (\delta u^+ - \delta d^+)}{5 (2 g^u_V - g^d_V) (u^+ + d^+)}\, ,\nonumber\\
\Delta^s_{AV} &=& - \frac{4 (g^u_V + 2 g^d_V) s^+}{5 (2 g^u_V - g^d_V) (u^+ + d^+)}\, ,\nonumber\\
\Delta^c_{AV} &=& \frac{4 (g^u_V + 2 g^d_V) c^+}{5 (2 g^u_V - g^d_V) (u^+ + d^+)}\, ,
\eea
and
\bea
\label{eq:Delta-VA}
\Delta^{\rm CSV}_{VA} &=& \frac{- 2 g^u_A \delta d_V + g^d_A \delta u_V}{(2 g^u_A - g^d_A) (u_V + d_V)} 
+ \frac{4 \delta d^+ + \delta u^+}{5(u^+ + d^+)}\, ,\nonumber\\
\Delta^s_{VA} &=& - \frac{2 g^d_A s_V}{(2 g^u_A - g^d_A) (u_V + d_V)} - \frac{2 s^+}{5(u^+ + d^+)}\, ,\nonumber\\
\Delta^c_{VA} &=& \frac{4 g^u_A c_V}{(2 g^u_A - g^d_A) (u_V + d_V)} - \frac{8 c^+}{5(u^+ + d^+)}\, .
\eea
In addition, an overall correction factor can be defined,
\be
A^{e^-}_{RL,d} = \frac{3 G_F Q^2}{10 \sqrt{2} \pi \alpha} 
\Big[ (2 g^{eu}_{AV} - g^{ed}_{AV})  
+ R_V Y (2 g^{eu}_{VA} - g^{ed}_{VA}) \Big]
(1 + \Delta^{e^-})\, ,
\ee
where
\be
\label{eq:Detla_elec}
\Delta^{e^-}
= \frac{(2 g^{eu}_{AV} - g^{ed}_{AV}) \Delta_{AV}  
+ R_V Y (2 g^{eu}_{VA} - g^{ed}_{VA}) \Delta_{VA}}{(2 g^{eu}_{AV} - g^{ed}_{AV})  
+ R_V Y (2 g^{eu}_{VA} - g^{ed}_{VA})}\, .
\ee
The individual contributions to $\Delta^{e^-}$ from CSV, $s$ and $c$ quark distributions can be obtained accordingly.

%%%%%%%%%%%%%%%%%%%%%%%%%%%%%%%%%%%%%%%%%%%%%%
\subsection{$A^{e^+ e^-}_d$}
The asymmetry between unpolarized $e^+$ and $e^-$ beams scattering from a deuteron target is
\be
A^{e^+ e^-}_{d} = - \frac{3 G_F Q^2}{2 \sqrt{2} \pi \alpha} \frac{1- (1-y)^2}{1 + (1-y)^2} \frac{R_V (2 g^{eu}_{AA} - g^{ed}_{AA})}{5 + 4 R_c + R_s} (1 + \Delta^{e^+ e^-})\, .
\ee
The correction term, $\Delta^{e^+ e^-}$, can be written as
\be
\Delta^{e^+ e^-} = \Delta_{\rm CSV}^{e^+ e^-} + \Delta_s^{e^+ e^-} + \Delta_c^{e^+ e^-}\, ,
\ee
where
\bea
\Delta^{e^+ e^-}_{\rm CSV} &=& \frac{- 2 g^u_A \delta d_V + g^d_A \delta u_V}{(2 g^u_A - g^d_A) (u_V + d_V)} 
+ \frac{4 \delta d^+ + \delta u^+}{5(u^+ + d^+) + 8 c^+ + 2 s^+}\, ,\nonumber\\
\Delta^{e^+ e^-}_s &=& - \frac{2 g^d_A s_V}{(2 g^u_A - g^d_A) (u_V + d_V)}\, ,\nonumber\\
\Delta^{e^+ e^-}_c &=& \frac{4 g^u_A c_V}{(2 g^u_A - g^d_A) (u_V + d_V)}\, .
\eea
%

%%%%%%%%%%%%%%%%%%%%%%%%%%%%%%%%%
\section{Numerical results}
\label{sec:results}
These correction factors depend on the weak couplings given in~(\ref{eq:couplings}) which are defined by $\sin^2\theta_W$.
In our analysis, we take the SM value of $\sin^2\theta_W$ in the $\overline{\rm MS}$ scheme, 
$\sin^2\theta_W(\mu = 0)|_{\overline{\rm MS}} = 0.23863 \pm 0.00005$~\cite{ParticleDataGroup:2022pth}.

\subsection{NNPDF}
In order to estimate the size of the contributions from CSV and the heavy quark PDFs, we first take the latest PDF set at NNLO from the NNPDF Collaboration~\cite{NNPDF:2023tyk}. There all PDFs, including $s - \bar{s}$ and $c - \bar{c}$, were determined self-consistently in the $\overline{\rm MS}$ scheme. We have converted these PDFs to the DIS scheme using the appropriate coefficient functions~\cite{Collins}.
The CSV distributions $\delta u$ and $\delta d$ are taken from~\cite{Wang:2015msk}, which combined the effects of both the $u$-$d$ mass difference and QED. 
Here we will impose $30\%$ uncertainties on $\delta u$ and $\delta d$ corresponding to the uncertainties in the lattice determination of their second moments~\cite{Shanahan:2013vla}.

\begin{figure*}[!h]
\begin{center}
\includegraphics[width=\columnwidth]{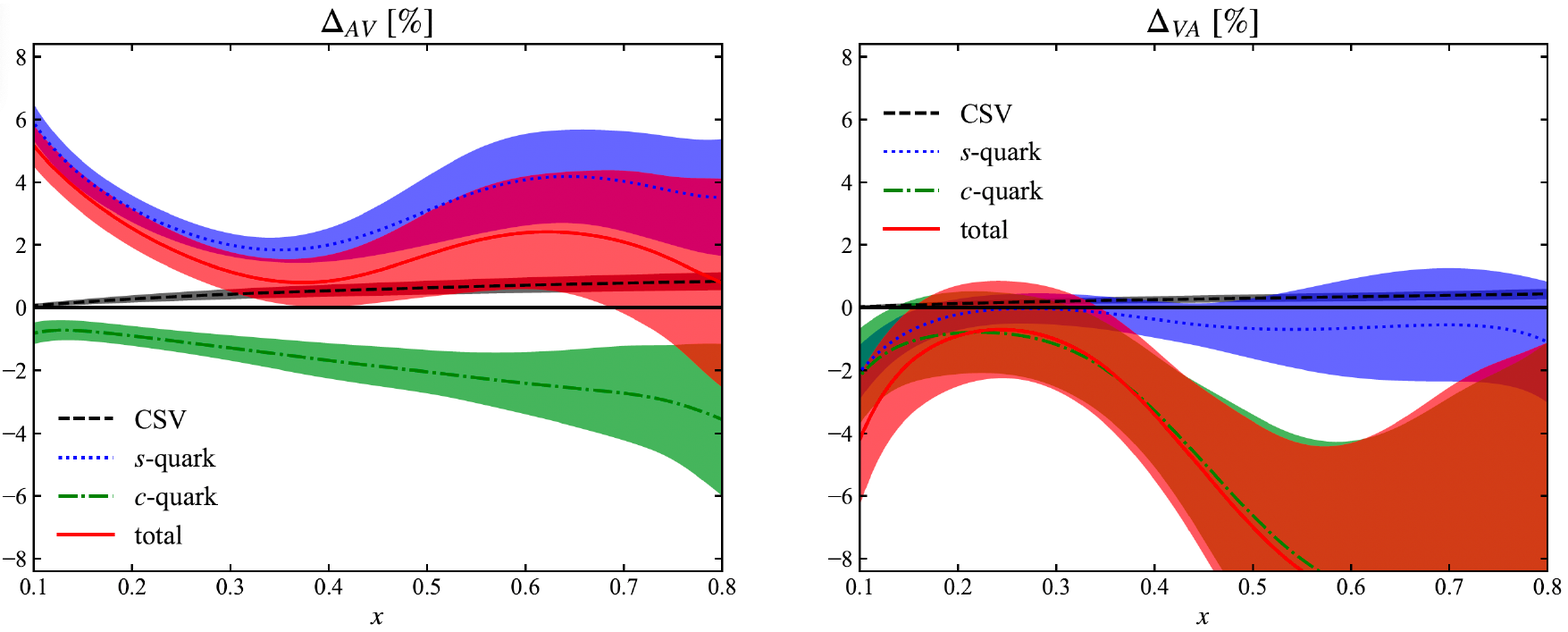}
\vspace*{-0.2cm}
\caption{Correction factors $\Delta_{AV}$ and $\Delta_{VA}$ at $Q^2 = 10\ {\rm GeV}^2$ (in percentage).}
\label{fig:Delta_C1q_C2q_10GeV2_nnpdf}
\end{center}
\end{figure*}

The correction factors $\Delta_{AV}$ and $\Delta_{VA}$ are shown in figure~\ref{fig:Delta_C1q_C2q_10GeV2_nnpdf}. We stress that the former involve only the charge conjugation positive PDFs, $s^+$ and $c^+$, while the latter also involve the less well known C-odd distributions $s_v$ and $c_v$. The error bands are associated with the uncertainties in the $s$ and $c$ quark distributions. 

There is strong cancellation between the $s$-quark and $c-$quark contributions in $\Delta_{AV}$, leading to a total $1\% - 2\%$ correction in the valence region.
This means the extracted value of $g^{eq}_{AV}$ could be overestimated by $(1- 2)\%$ if these corrections were neglected. In addition, because of the cancellation between the effects of $s^+$ and $c^+$, the total correction could be larger if the NNPDF result for one of these flavors were to be modified by later analysis. The correction associated with CSV is comparatively much smaller.

Once again, for $\Delta_{VA}$ the effect of CSV is quite small. In contrast with the results for $\Delta_{AV}$, this correction term is dominated by the $c-$quark contribution, while the $s-$quark and CSV effects play a less important role. 
While the contributions from the C-odd and C-even combinations tend to cancel, the coefficient of $c^+$ in the expression for $\Delta^c_{VA}$ 
( c.f.~(\ref{eq:Delta-VA}) is so large that the charm quark contribution dominates in this case. Neglecting these corrections would result in underestimating the coupling $g^{eq}_{VA}$. However, the most important lesson is that given the tremendous uncertainty in the C-odd PDFs discussed earlier, the actual corrections may well be even larger than indicated here.  

\begin{figure}[!h]
\begin{center}
\includegraphics[width=0.8\columnwidth]{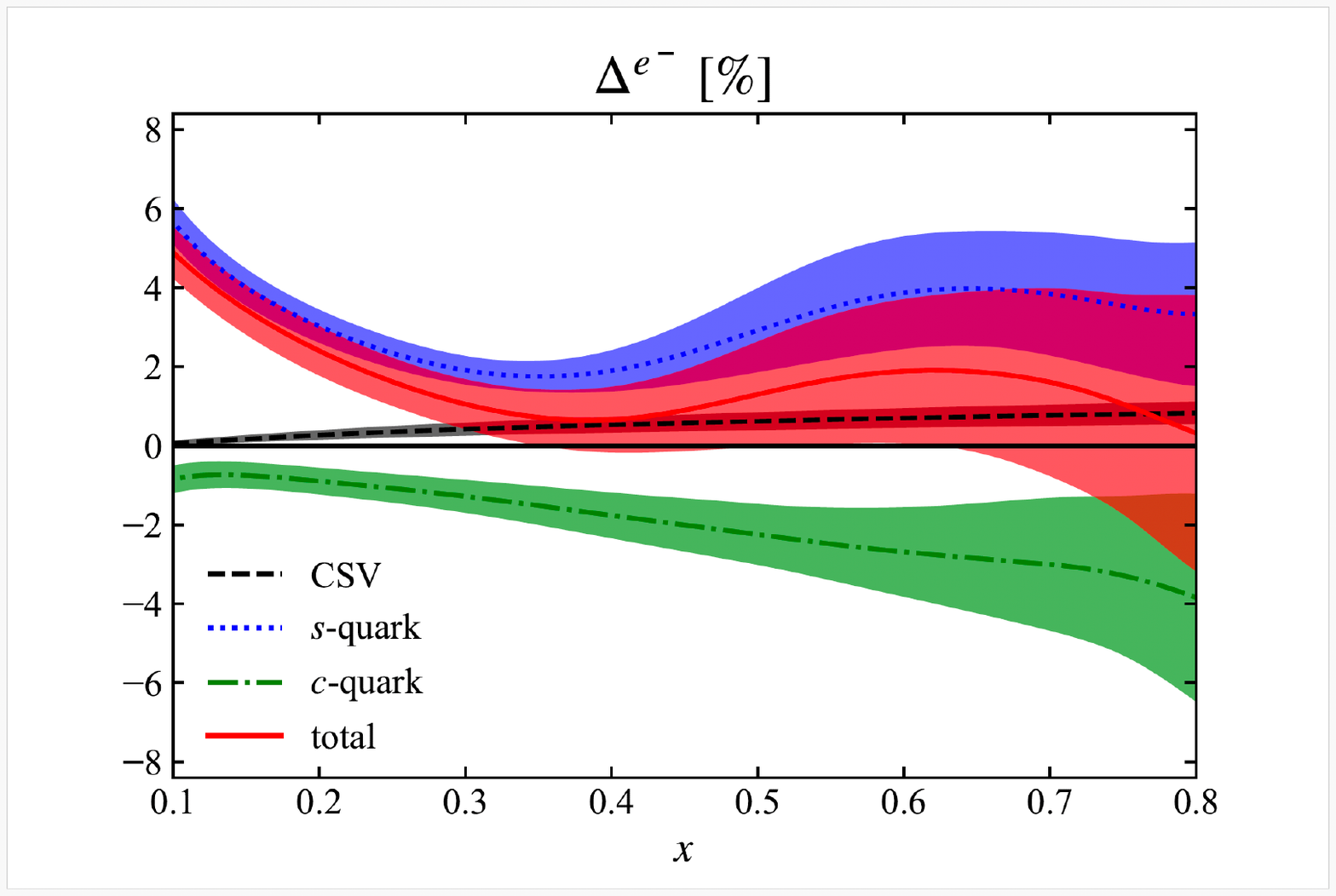}
\vspace*{-0.2cm}
\caption{Corrections to $A^{e^-}_{RL,d}$ (in percentage) at $Q^2 = 10\ {\rm GeV}^2$ for $y=0.5$.}
\label{fig:Delta_elec_10GeV2_nnpdf}
\end{center}
\end{figure}

The overall correction to $A^{e^-}_{RL,d}$ is shown in figure~\ref{fig:Delta_elec_10GeV2_nnpdf}, taking $y=0.5$ as an example.
Note that the $\Delta_{AV}$ term is dominant in~(\ref{eq:Detla_elec}), since its coefficient is much larger than that of $\Delta_{VA}$.


\begin{figure}[!h]
\begin{center}
\includegraphics[width=0.8\columnwidth]{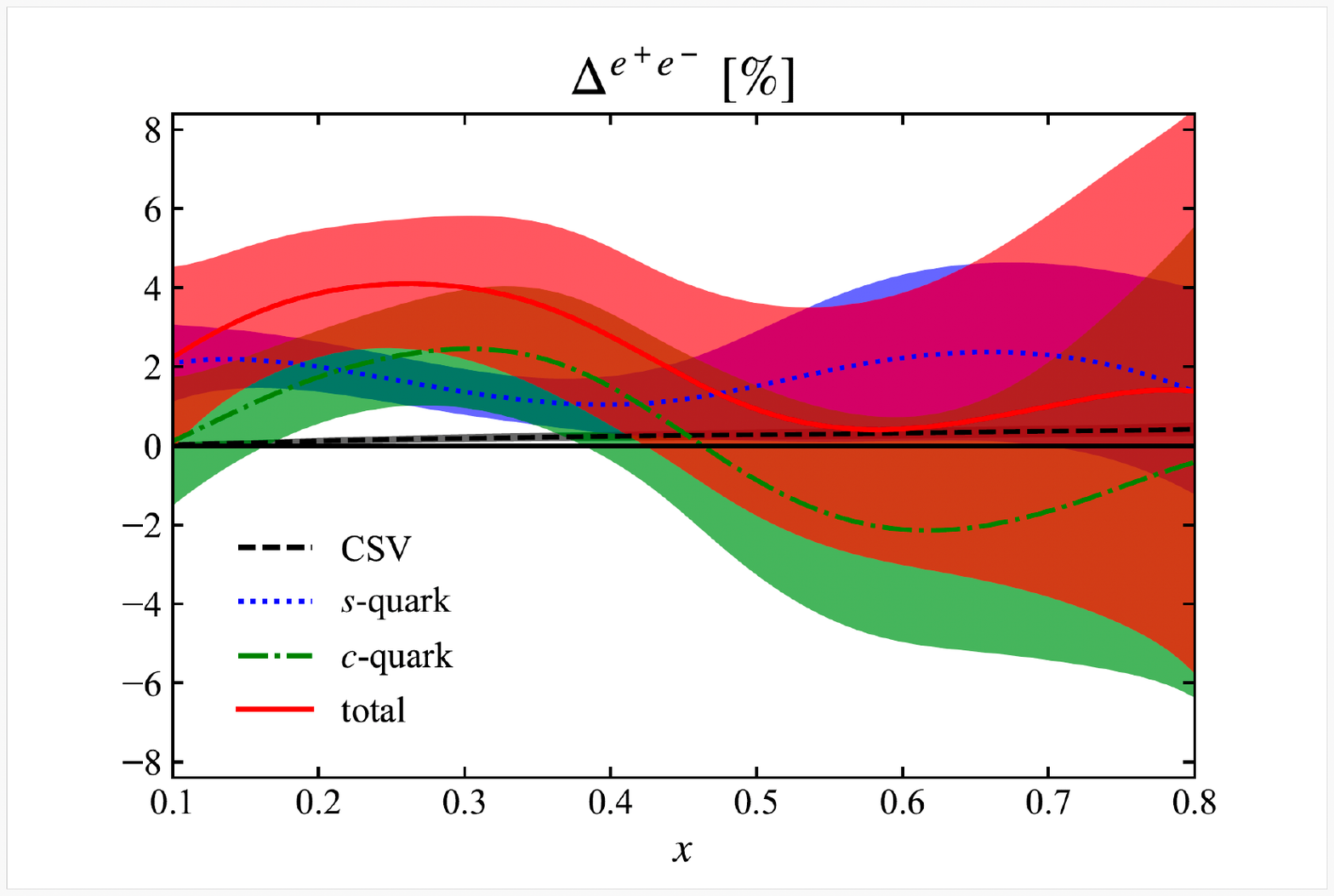}
\vspace*{-0.2cm}
\caption{Corrections to $A^{e^+ e^-}_{d}$ (in percentage) at $Q^2 = 10\ {\rm GeV}^2$.}
\label{fig:Delta_C3q_10GeV2_nnpdf}
\end{center}
\end{figure}

The correction to $A^{e^+ e^-}_{d}$ is given in figure~\ref{fig:Delta_C3q_10GeV2_nnpdf}. In the region 
$x < 0.5$, $\Delta^{e^+ e^-}$ is dominated by the sum of the $s$ and $c$ quark contributions, amounting to $4\%$ or more. For $x > 0.5$, there is strong cancellation between the $s$ and $c$ quark contributions, leading to a total $(1-2)\%$ correction to $A^{e^+ e^-}_{d}$.
These corrections are subject to large uncertainties, originating from those in the strange and charm quark distributions.

\subsection{CT18 and MSHT20}

\begin{figure}[!h]
\begin{center}
\includegraphics[width=\columnwidth]{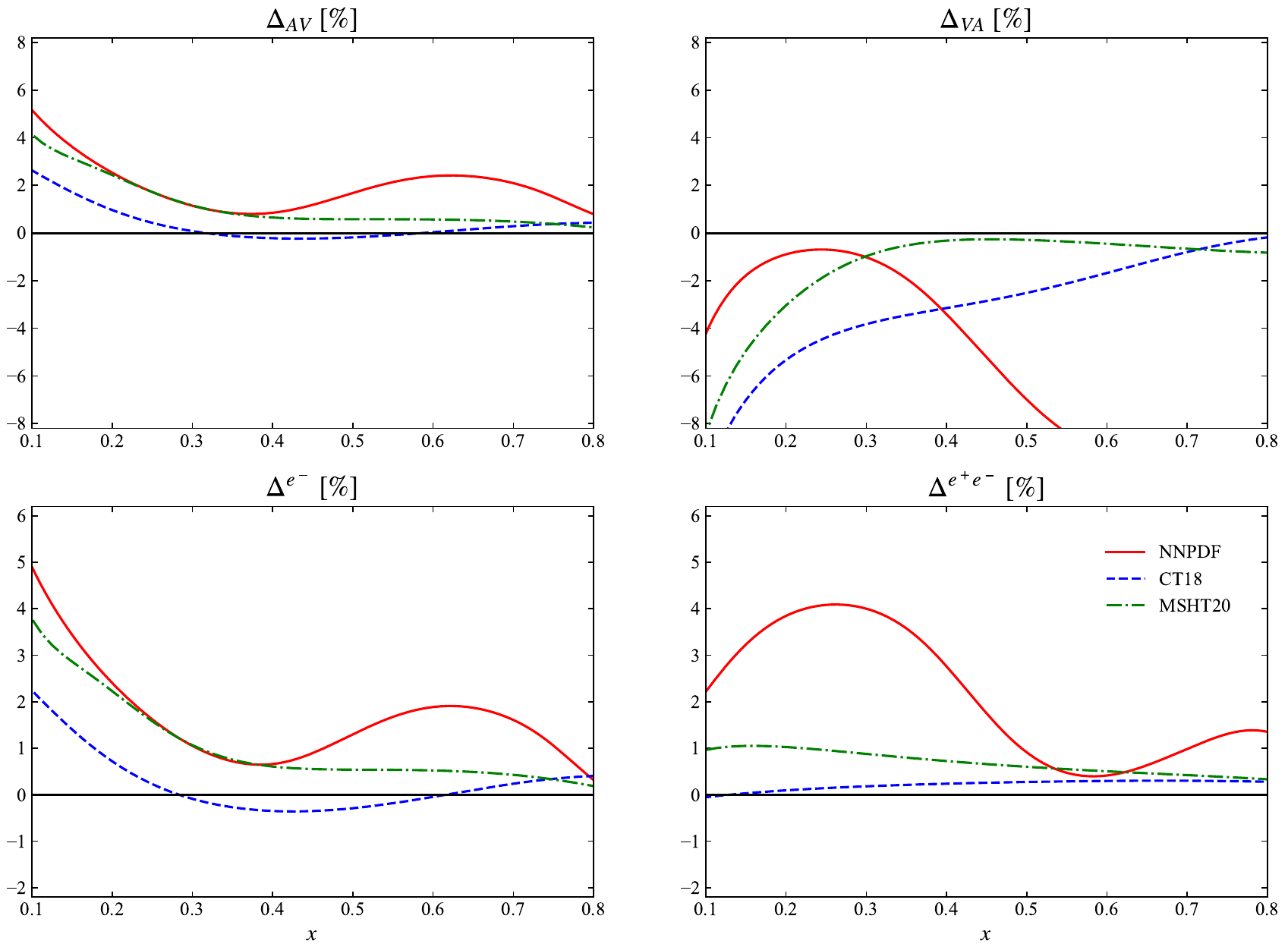}
\vspace*{-0.2cm}
\caption{Comparison of the correction factors using NNPDF, CT18, and MSHT20 NNLO PDF sets at $Q^2 = 10\ {\rm GeV}^2$.}
\label{fig:comparison}
\end{center}
\end{figure}

As we mentioned earlier, the discrepancies in the strange and charm quark distributions among different global fit analyses may be even larger than the PDF uncertainties. Therefore, we also calculate the correction factors using the CT18 NNLO~\cite{Guzzi:2022rca} and MSHT20 NNLO~\cite{Bailey:2020ooq} PDF sets. The central values of the total contributions are compared with the NNPDF results as shown in figure~\ref{fig:comparison}. The discrepancies constitute even larger systematic uncertainties on these correction factors.

\section{New physics searches}
\label{sec:new-physics}

Although the correction factors are relatively small and may be consistent with zero in some cases, the current large uncertainties would have a significant impact on new physics searches beyond the SM.

Following the convention in~\cite{Erler:2014fqa}, new contact interaction terms with coefficients $g^2/\Lambda^2$, characterising physics beyond Standard Model, can be added to the neutral-current SM Lagrangian. The new physics is usually considered to be strongly coupled with $g^2 = 4\pi$~\cite{Eichten:1983hw}. The sensitivity of PVES experiments to new physics is thus set by the energy scale $\Lambda$.

Elastic PVES experiments off a proton target provide access to the weak couplings $g_{AV}^{eq}$ through measurements of the proton weak charge. The Qweak experiment set an upper limit of $\Lambda = 26.3\ {\rm TeV}$~\cite{Qweak:2018tjf}, while the proposed P2 experiment with improved precision of $1.7\%$ could reach a new physics scale of order $50\ {\rm TeV}$~\cite{Becker:2018ggl, Erler:2014fqa}.

The SoLID experiments are designed to extract the couplings $g_{VA}^{eq}$ by measuring the parity violation in deep inelastic scatterings with an expected precision of order $0.6\%$, corresponding to the mass reach $\Lambda=22\ {\rm TeV}$ at $95\%$ CL~\cite{Erler:2014fqa}. Moreover, SoLID has the unique feature of providing the first measurement of the coupling $g^{eq}_{AA}$~\cite{Arrington:2021alx}.

To explore the impact of these corrections on constraining physics beyond the Standard Model, we estimate the new physics scale $\Lambda$ as in~\cite{Zheng:2021hcf}.
The maximal $1\sigma$-sensitivity is given by
\be
\Lambda = v \sqrt{\frac{8\sqrt{5} \pi}{\Delta (2 g^{eu}_{AA} - g^{ed}_{AA})}}\, ,
\ee
where $v = (\sqrt{2} G_F)^{-1/2} = 246.22\ {\rm GeV}$ is the Higgs vacuum expectation value. A targeted precision of $\Delta (2 g^{eu}_{AA} - g^{ed}_{AA}) = \pm 0.03$, which amounts to a $2\%$ uncertainty, would imply $\Lambda = 10.7\ {\rm TeV}$~\cite{Zheng:2021hcf}.
However, an extra $4\%$ systematic uncertainty from the correction factor $\Delta^{e^+ e^-}$ presented in this work could significantly reduce the new physics reach to $\Lambda = 6.2\ {\rm TeV}$.

Large systematic uncertainties on these weak couplings could also mimic potential new physics effects, such as a dark photon, which could potentially lead to a shift in $g^{eq}_{AA}$ as large as $5\%$ at $Q^2 = 10\ {\rm GeV}^2$~\cite{Thomas:2022qhj}.

%%%%%%%%%%%%%%%%%%%XXXXXXXXXXXXXXXXXXXXXXX
\section{Conclusion}
\label{sec:conclusion}
We have derived the heavy quark and charge symmetry violation corrections to the parity-violating asymmetry and lepton-charge asymmetry in deep inelastic scattering from a deuteron target.

Based on the latest PDF set from the NNPDF Collaboration, the heavy quark corrections relevant to $g^{eq}_{AV}$ and $g^{eq}_{VA}$ are of order $(1-2)\%$ at $Q^2 = 10\ {\rm GeV}^2$, but opposite in sign. The correction to the asymmetry $A^{e^+ e^-}_d$ implies that the uncertainty in $g^{eq}_{AA}$ extracted from the data could be of order $4\%$ or larger.
We have also checked the results using the CT18 and MSHT20 NNLO PDF sets, which could result in even larger systematic uncertainties.

These corrections must be taken into account for a precise determination of the weak couplings and new physics searches from electron scattering experiments. 
Moreover, our current knowledge of the strange and charm quark distributions is still poor due to large uncertainties and inconsistency among different global fit analyses, undermining the potential of PVES to have its maximum impact on the extraction of new physics beyond the SM.
It is therefore vital that there should be a coherent program of high precision measurements, which provide information not only concerning  Standard Model parameters but also on charge symmetry violation in the valence PDFs and the C-even and C-odd distributions of strange and charm quarks.

%%%%%%%%%%%%%%%%%%%%%%%%%%%%%%%%%%%%%%%%%%%%%%
\ack{
We are pleased to acknowledge helpful discussions with Xiaochao Zheng.
This work was supported by the University of Adelaide and the Australian Research Council through the Centre of Excellence for Dark Matter Particle Physics (CE200100008).
}
%%%%%%%%%%%%%%%%%%%%%%%%%%%%%%%%%%%%%%%%%%%%%%%%%%%%%%%

\end{document}